\documentclass[fleqn,usenatbib]{mnras}

\usepackage{newtxtext,newtxmath}
\usepackage{pdflscape}
\usepackage{tabularx}
\usepackage{longtable}
\usepackage{xcolor,colortbl}
\definecolor{aqua}{rgb}{0.0, 1.0, 1.0}
\definecolor{amber}{rgb}{1.0, 0.75, 0.0}
\definecolor{americanrose}{rgb}{1.0, 0.01, 0.24}
\definecolor{amethyst}{rgb}{0.6, 0.4, 0.8}
\definecolor{cambridgeblue}{rgb}{0.64, 0.76, 0.68}
\definecolor{brightgreen}{rgb}{0.4, 1.0, 0.0}
\usepackage{rotating}

\usepackage{adjustbox}
\usepackage[T1]{fontenc}
\usepackage{threeparttable}
\DeclareRobustCommand{\VAN}[3]{#2}
\let\VANthebibliography\thebibliography
\def\thebibliography{\DeclareRobustCommand{\VAN}[3]{##3}\VANthebibliography}

\usepackage{graphicx}	
\usepackage{amsmath}	
\usepackage{xcolor}

\title[Spectral Obs. of PNe from the the HASH Database]{Spectroscopic Observations of Selected Planetary Nebulae from the HASH Database}

\author[Temiz et al.]{
Utkan Temiz$^{1}$,
Nazım Aksaker$^{2,1}$\thanks{E-mail: naksaker@cu.edu.tr}
and 
Aysun Akyuz$^{3,1}$
\\
$^{1}$Space Science and Solar Energy Research and Application Center (UZAYMER), University of Çukurova, 01330, Adana, Turkey\\
$^{2}$Adana Organised Industrial Zones Vocational School of Technical Science, Çukurova University, 01410, Adana, Turkey\\
$^{3}$Department of Physics, University of Çukurova, 01330, Adana, Turkey\\
}
\date{Accepted 2023 September 13. Received 2023 August 31; in original form 2023 May 31}

\pubyear{2023}

\begin{document}
\label{firstpage}
\pagerange{\pageref{firstpage}--\pageref{lastpage}}
\maketitle

\begin{abstract}
We conducted research on the classification and physical properties of 10 objects from the HASH (Hong Kong/Australian Astronomical Observatory/Strasbourg Observatory H-alpha Planetary Nebula (PN)) database with small angular sizes (< 8\arcsec) in the northern hemisphere. The sample consisted of 6 Likely PNe, 2 new candidates, one emission-line star, and one object of unknown nature. Among them, we observed 4 objects for the first time using the medium-resolution TFOSC spectrograph located on the RTT150 cm of the TÜBİTAK National Observatory (TUG). To investigate the classification of the observed objects, we utilized the emission line ratios of [O III]/H$_{\gamma}$, [O III]/H$_{\beta}$, [N II]/H$_{\alpha}$ and [S II]/H$_{\alpha}$ and diagnostic diagrams such as the Sabbadin-Minello-Bianchini (SMB) and Baldwin-Phillips-Terlevich (BPT). When considering a broader range of diagnostic criteria compared to those provided in the literature, our analyses resulted in the reclassification of 4 objects from Likely PNe to True PNe and the retention of the previous classification for the remaining 6 objects. In addition, we obtained various physical conditions such as electron temperatures, electron densities, logarithmic extinction coefficients, and excitation classes for the 10 objects under study. Our analysis revealed that the ionic abundances of the majority of these objects were in agreement with Galactic PNe. Our spectral observations have led to the updating of 10 PNe in the HASH database.
\end{abstract}




\section{Introduction}
Planetary Nebulae (PNe) are the final stage in the evolution of stars with masses ranging from $\sim$ 0.8 to 8 M\sun. They play a crucial role in enriching the chemical composition of the interstellar medium through the ejection of elements from their progenitor stars. Furthermore, as a source of ionizing radiation, PNe have an impact on the ionization state and emission characteristics of the interstellar medium \citep{2006agna.book.....O}. The study of PNe spectra, which exhibit bright optical emission lines, is an essential tool for determining the physical properties of the ionized gas \citep{2022PASP..134b2001K}. These spectral features arise mostly from collisionally excited emission lines of ions such as Oxygen, Nitrogen, Neon, Argon, and Sulphur, as well as standard Hydrogen and Helium recombination lines. Accurate measurements of these spectral signatures enable the calculation of ionic abundances and physical conditions in the ionized gas, providing basic insights into the properties and evolutionary history of PNe (for recent reviews, see \citealp{2022FrASS...9.5287P, 2022PASP..134b2001K}).

Since the first observation of a PN (M27; dumbbell nebula) by Charles Messier in 1764, thousands of PNe have been observed in the Milky Way and numerous catalogs and surveys have been published, including those by \cite{1967BAICz..18..252P,1992secg.book.....A,1996fsse.book.....A, 2001A&A...378..843K, 2008MNRAS.384..525M, 2005MNRAS.362..753D, 2014MNRAS.440.2036D, 2014MNRAS.443.3388S}. Currently, there is a comprehensive multiwavelength database known as the Hong Kong/Australian Astronomical Observatory/Strasbourg Observatory H-alpha Planetary Nebula (HASH) Database\footnote{http://202.189.117.101:8999/gpne/index.php}. This interactive database offers access to a wide range of data, including imaging, spectroscopic, and other observational data related principally to Galactic PNe \citep{2016JPhCS.728c2008P}. The presence of this database has greatly contributed to the advancements in the field of PNe research. 
 
In various studies on PNe in the literature, HASH has been used to determine their angular sizes \citep{2021MNRAS.503.2887B}, identify central stars \citep{2023MNRAS.520..773R, 2022Galax..10...32P}, and analyze their morphology \citep{2023MNRAS.519.1049T}. Additionally, a recent study by \cite{2023NewA...9801943P} confirmed that an object previously classified as an emission line star in the HASH database as part of an amateur search for new PN candidates is a symbiotic star whose optical spectrum shows distinct emission lines. 

The HASH database has categorized a total of 10110 objects up to date of 07.17.2023 with the following classifications: True PNe (2705), Possible PNe (710), Likely PNe (465), New Candidates (2432), and other objects sharing similarities with PNe (3798).
As it is well known, PNe displays a diverse range of observational and physical properties across a wide range of wavelengths, spanning from ultraviolet to radio regimes \citep{2019MNRAS.488.3238A, 2022FrASS...9.5287P, 2023arXiv230401970G}. Optical narrow-band images, along with high-quality multi-wavelength data, are particularly valuable in the identification of PNe candidates and differentiating them from their mimics. However, an optical spectrum plays a crucial role in achieving a comprehensive diagnostic identification \citep{2010PASA...27..129F,2022FrASS...9.5287P}. Numerous observation programs have been implemented to both discover and confirm Galactic PNe using spectroscopic data obtained from ground-based telescopes. Some of these programs have greatly benefited from the valuable contributions made by amateur observers \citep{2012RMxAA..48..223A,2022A&A...666A.152L, 2023MNRAS.520..773R}.

In this study, we aim to reclassify and determine the physical properties of 10 objects (6 Likely PNe, 2 new candidates, one emission-line star, and one unknown nature) that were selected from the HASH and observed as part of a spectroscopic observation program conducted at the TÜBİTAK National Observatory ({\it TUG}\footnote{https://tug.tubitak.gov.tr}). We noted that these objects have spectra in the HASH, but those suffer from a low signal-to-noise ratio and low spectral resolution. We, therefore, conducted new higher-quality observations. Selection criteria were applied to choose these 10 objects from the HASH database, including being observable from the northern sky (Dec. >-10$^{\circ}$), having sizes smaller than 8\arcsec (with three exceptions), and lacking detailed studies in the literature. Notably, spectral observations of four out of the 10 selected objects were conducted for the first time. However, the remaining 6 objects were further examined to provide more precise confirmation and investigate their physical conditions, despite having spectral observations available in the HASH database or the literature \citep{1984ApJ...278..610B, 1992secg.book.....A, 1996PASP..108..980K, 2010A&A...511A..44S}.

This paper is organized as follows: a description of spectral observations and the data reduction are presented in Section \ref{sec:obs}. The physical conditions and ionic abundances of PNe and diagnosis are given in Section \ref{sec:obs}. 
We present a summary and discussion of the obtained results
in Section \ref{sec:res}. Finally, our main findings and classification results are presented in Section \ref{sec:con}.

\section{Observations and Data Analysis}
\label{sec:obs}
The optical spectral data of 10 selected objects were obtained from the TUG in Bakırlıtepe-Antalya/Turkey. {\it TUG} Faint Object Spectrograph and Camera (TFOSC) mounted on the Russian-Turkish Telescope 1.5 m (RTT150) Cassegrain focus was used to provide spectroscopic and imaging data. The observations were carried out during the period from 2017 to 2020. During this period, three different CCD cameras (Fairchild 447, Andor DW436-BV, and Andor iKon-L 936 BEX2-DD-9ZQ) were used for observations. All CCD chips have the same element of pixels as 2048x2048. The pixel scales are 0\farcs39/pixel and 0\farcs33/pixel for Fairchild 447 and Andor CCDs, respectively. In these observations, the spectra of the objects were taken with two different grisms in accordance with our goal.
These are, grism 8 (6200-7850 {\AA}, $\Delta$$\lambda$ 3.0 {\AA}), grism 14 (3270-6120 {\AA}, $\Delta$$\lambda$ 5.4 {\AA}), which cover the red and blue region, respectively. Three slit units of 1\arcsec.78 x 676\arcsec, 2\arcsec.38 x 234\arcsec and 3\arcsec.92 x 676\arcsec were used depending on the size of the objects and the seeing conditions.
The exposure time for each grism consisted of 3600s for spectral observations.
A detailed list of objects is presented in Table \ref{T:obslog}, which includes their coordinate information, major diameters, main class obtained from the HASH database, {\it R} magnitudes, distances, and an observational log containing dates, grisms, and seeing conditions. In our list, there are 6 Likely PNe, 2 new candidates, one emission-line star, and one object of unknown nature. According to the ERBIAS scheme described in \cite{2020arXiv201205621P}, the main classes assigned to these objects are elliptical (E), spherical (S), and bipolar (B). All objects in the list are within TFOSC's red (R) magnitude (R$_{max}$ $\leq$ 22.4).

\begin{figure*}
\begin{center}
\includegraphics[angle=0,scale=0.35]{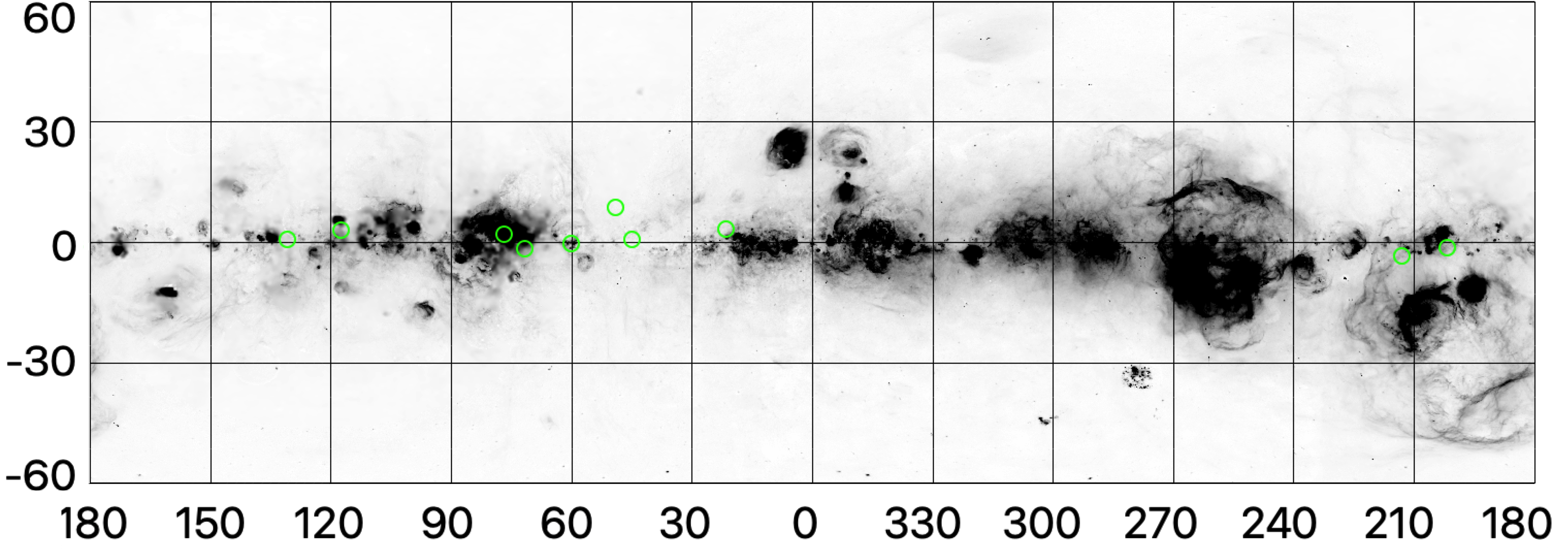}
\caption{The Galactic distribution of objects (green circles; sizes are not scaled) selected from HASH are shown on the H$_{\alpha}$ map given by \protect\cite{2003ApJS..146..407F}. 
The image represents an inverted-color map where dark regions indicate the presence of H II regions. The objects are distributed along the Galactic plane with regions between Orion's arm ({\it l}=220$^{\circ}$ and {\it b}=-15$^{\circ}$) and the Galactic center ({\it l}=0$^{\circ}$ and {\it b}=0$^{\circ}$) being outside our field of view. Axes units are in degrees. The X and Y axes represent Galactic longitude ({\it l}) and latitude ({\it b}), respectively.} 
\label{F:sky}
\end{center}
\end{figure*}

It is noteworthy that when examining the spectrum of a PN, the H$_{\alpha}$ line is observed as a significant emission line. Therefore, for comparative purposes, we have plotted the coordinates of our objects on the H$_{\alpha}$ map of the Galaxy, as shown in Fig. \ref{F:sky}. Notably, the longitudes of our objects span from 20$^{\circ}$ to 220$^{\circ}$, covering almost two-thirds of the Galactic plane.

\begin{figure*}
\begin{center}
\begin{tabular}{ |c|c|c|c|c|c| }
\hline
021.2+02.9&044.6+00.4&048.5+04.2&059.8-00.5&071.4-01.9\\
\hline
    \includegraphics[angle=0,scale=0.19]{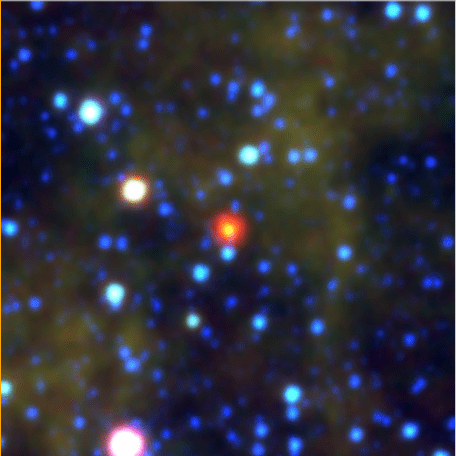}&
    \includegraphics[angle=0,scale=0.19]{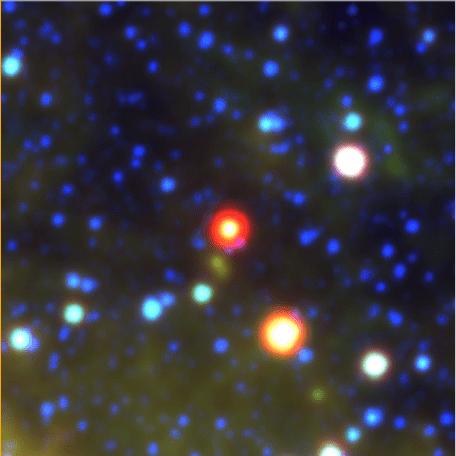}&
    \includegraphics[angle=0,scale=0.19]{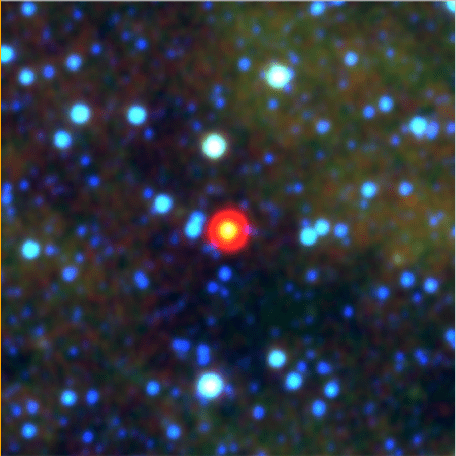}&
    \includegraphics[angle=0,scale=0.19]{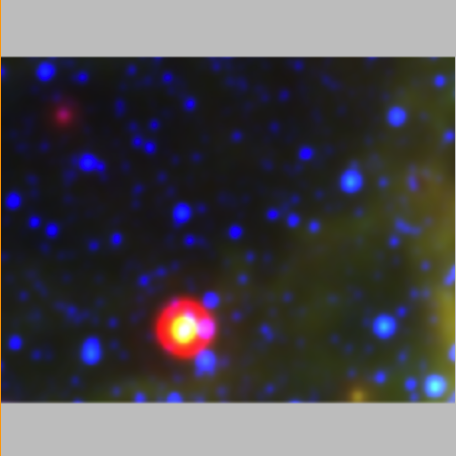}&
    \includegraphics[angle=0,scale=0.19]{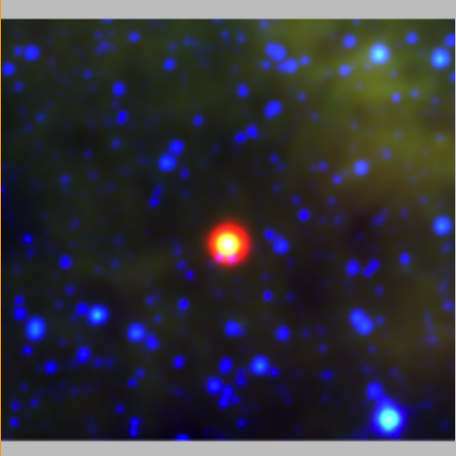}\\
\hline   
076.4+01.8&117.2+02.6&130.4+00.4&201.5-01.6&212.8-03.6\\
\hline 
    \includegraphics[angle=0,scale=0.19]{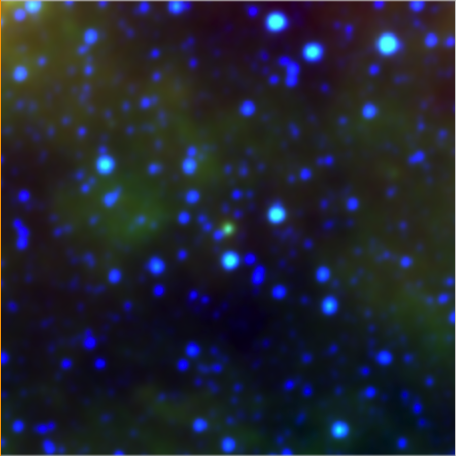}&
    \includegraphics[angle=0,scale=0.19]{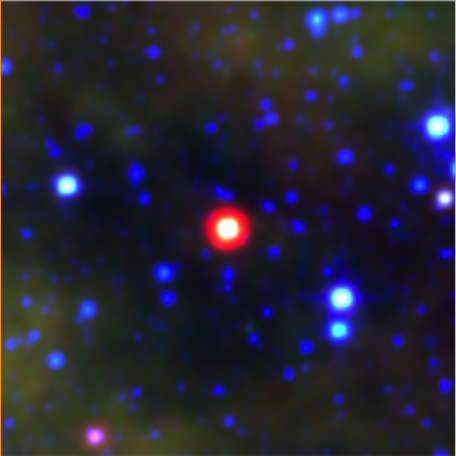}&
    \includegraphics[angle=0,scale=0.19]{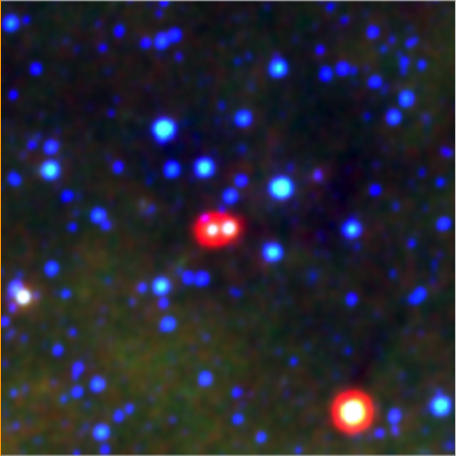}&
    \includegraphics[angle=0,scale=0.19]{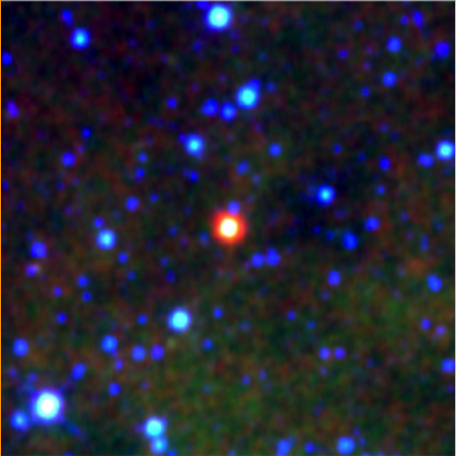}&
    \includegraphics[angle=0,scale=0.19]{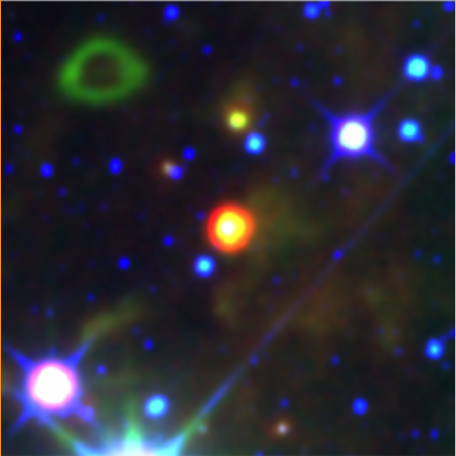}\\
    \hline
\end{tabular}
\caption{RGB (Red:W4, Green:W3 and Blue:W2) of WISE images of 10 objects selected from HASH. Images are ordered by the name of PN G designations as given in Table \ref{T:obslog}. The objects are reddish in color at the center of the images except for PN G076.4+01.8. The size of the images is 10$\arcmin$ $\times$ 10$\arcmin$, except PN G059.8-00.5 and PN G071.4-01.9 which are not in the center because of the different sizes of RGB filters. For all images, the north is up and the east is left.} 
\label{F:WISE}
\end{center}
\end{figure*}

We also used WISE (Wide-field Infrared Survey \footnote{https://irsa.ipac.caltech.edu/}) images for visualization of these objects. The WISE mission has mapped the entire sky at four infrared wavelengths with much greater sensitivity than previous infrared missions. The filters corresponding to these four wavelengths are named W1 (3.4 $\mu$m), W2 (4.6 $\mu$m), W3 (12 $\mu$m), and W4 (22 $\mu$m), respectively.
We used only three WISE filters to generate RGB (Red:W4, Green:W3, and Blue:W2) images of the selected 10 objects shown in Fig. \ref{F:WISE}. 

We carried out comprehensive spectral analyses of the optical data of these objects. Standard data reduction steps were applied using the {\scshape iraf v 2.16} for spectral analyses. 
Initially, {\it zerocombine} and {\it ccdproc} tasks were used for bias subtraction and flat-fielding, respectively.
Then, {\it identify}, {\it reidentify} and {\it fitcoords} tasks were also used for wavelength calibration of spectra. Cosmic rays and hot pixels were removed using the {\it lacosmic} task. Then, 2D spectra were extracted and converted to 1D spectra with the {\it apall} task. 

Flux calibrations of the object spectra were performed with appropriate standard star observations selected from the Spectrophotometric Standards of {\href{https://www.eso.org/sci/observing/tools/standards/spectra/stanlis.html}
{ESO}}. These stars were \mbox{BD+25d4655}, \mbox{Feige 34}, \mbox{Feige 66}, and \mbox{BD+33d2642}, and the same data reduction steps were also applied to the spectra of these standard stars. Then, a flux calibration was performed with tasks {\it standard}, {\it sensitivity}, and {\it calibration} using observatory information, exposure time, and standard star information. Afterward, the task {\it wspec} was used to convert 1-D spectra to a text file thus the red and blue parts of spectra can be merged.

\begin{figure*}
\begin{center}
\includegraphics[angle=0,scale=0.5]{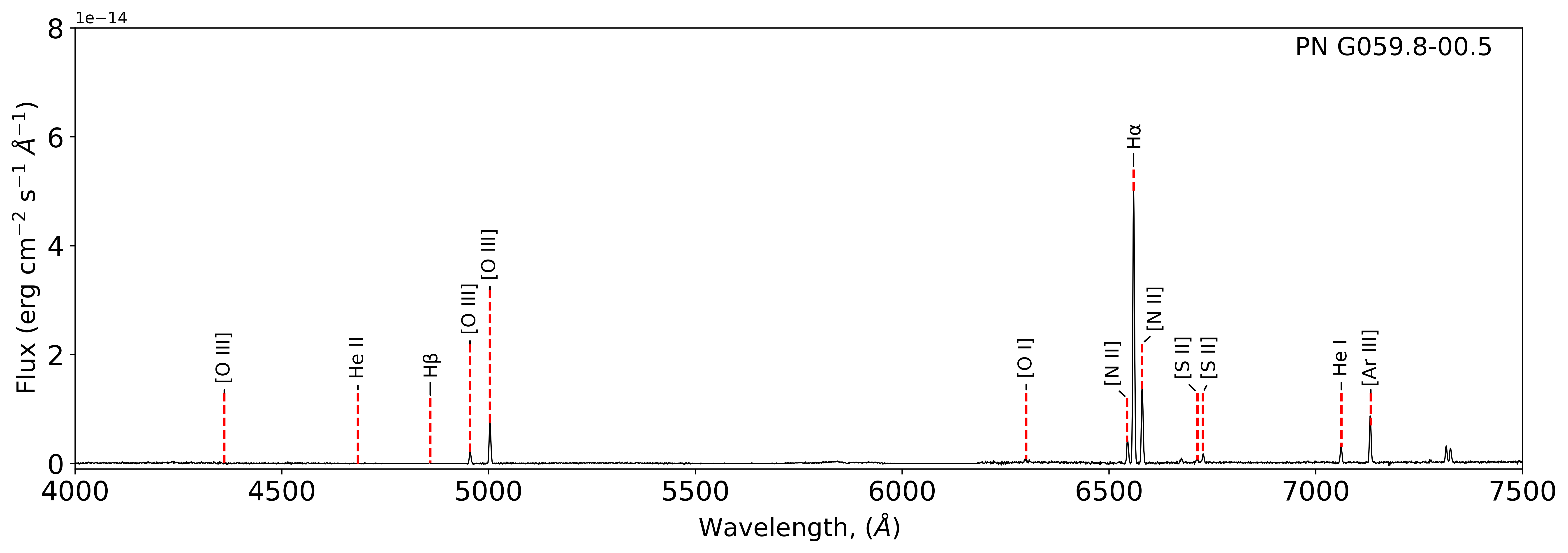}
\caption{The spectrum of PN G059.8-00.5 was observed for the first time and reclassified from Likely PN to True PN. X and y axes represent wavelength ({\AA}) and absolute flux (erg cm$^{-2}$ s$^{-1}$ {\AA}$^{-1}$), respectively. The wavelength range is 4000 - 7500 {\AA}. The spectrum combines grism 14 (blue part) and grism 8 (red part). The prominent emission features given in Table \ref{T:flux} are over-plotted.} 
\label{F:spec}
\end{center}
\end{figure*}

Most of the spectra obtained include forbidden lines of various ionization states expected from a typical ionized nebula, emission lines of recombination H and He transitions, and continuous emission. It is known that there are three different mechanisms to create the emission lines for photoionized nebulae. These are photoexcitation, recombination, and collisional excitation \citep{ 2017PASP..129h2001P}. Generally, for PNe, recombination and collisionally excited emission lines are used for determination and classification \citep{1981PASP...93....5B}. 
We measured emission line fluxes and their errors using the {\it splot} task in {\scshape iraf}. Emission line fluxes for each of the objects are given in Table \ref{T:flux} according to H$_{\beta} \lambda$4861 {\AA}. It is noteworthy that the ability to resolve the [N II] $\lambda$$\lambda$6548, 6584 {\AA} lines from the H$_{\alpha}$ line in the spectrum obtained with grism 8 demonstrates the resolving power of TFOSC. Our obtained spectra of the PNe are presented in Fig. \ref{F:spec} as an example, Fig. \ref{F:A1}, and Fig. \ref{F:A2}. The observed objects exhibit a minimum of 1 and a maximum of 11 identified emission lines in their spectra.

The spectral data of PNe are a valuable tool for calculating the H$_\beta$ logarithmic extinction coefficient, c(H$_\beta$), which quantifies the interstellar reddening around these objects \citep{2016MNRAS.455.1459F}. The formula used for calculating c(H$_\beta$) is 2.84*log((F(H$_{\alpha}$)/F(H$_{\beta}$))/2.86) \citep{2006agna.book.....O}.
We performed the reddening correction using the {\it deredden} task in {\scshape IRAF}. This task incorporated the empirical selective extinction function of Cardelli, Clayton, and Mathis (CCM89) \cite{1989ApJ...345..245C}. On the other hand, the reddening correction was not applied to PN G044.6+00.4, PN G071.4-01.9, and PN G130.4+00.4 due to the lack of H$_\beta$ emission line. Hence, we used the uncorrected reddening values for these sources.

As a subsequent procedure, we calculated specific emission line ratios including [O III]/H$_{\gamma}$, [O III]/H$_{\beta}$, [N II]/H$_{\alpha}$, and [S II]/H$_{\alpha}$. These ratios were used to create diagnostic diagrams, which are crucial in defining, classifying, and determining the physical conditions of the objects under investigation.

\subsection{Diagnosis}

Gaseous nebulae, such as H II regions, Supernova Remnants (SNRs), and PNe, display comparable emission lines in their optical spectra, although with varying intensities, due to their distinct emission mechanisms. However, diagnostic tools and emission line ratios have been employed to distinguish between these objects. One effective diagnostic for PNe involves the presence of the He II emission line at a wavelength of 4686 Å in the spectrum \citep{2010PASA...27..129F,2022FrASS...9.5287P}.

Notably, PNe are also distinguished by a strong [O III] $\lambda$5007 {\AA} line, making the [O III]/H$_{\beta}$ emission line ratio in the blue part of the spectrum as a diagnostic tool.  
This ratio also provides information about the excitation classes of the PNe \citep{2010PASA...27..129F}.
On the other hand, the [N II]/H$_{\alpha}$ ratio is another sensitive indicator
and its value can be influenced by the presence of shocks. If [N II]/H$_{\alpha}$ < 0.6, an object could be classified as either an H II region or a PN \citep{2010PASA...27..129F}.
The presence of shock waves in a nebula can lead to photoionization, resulting in the production of strong [S II] lines. 
An SNR can be identified by [S II]/H$_{\alpha}$ > 0.5 \citep{1985ApJ...292...29F}, while [S II]/H$_{\alpha}$ < 0.5 suggests that an object could be either a PN or an H II region \citep{1997ApJS..113..333M, 2010A&A...517A..91S}. We also calculate [O III]/H$_{\gamma}$ to construct a diagram ([O III] $\lambda$4363/H$_\gamma$ vs. [O III] $\lambda$5007/H$_\beta$) to help classify an object as either a PN or a Symbiotic Star. Detailed analysis is given in the section \ref{sec:res}.

We computed the emission line ratios including [O III]/H$_{\gamma}$, [O III]/H$_{\beta}$, [N II]/H$_{\alpha}$ and [S II]/H$_{\alpha}$ and their corresponding errors using the error propagation technique for our objects.
When these emission lines were present, their values were incorporated into Table \ref{T:Test}.
In this table, it should be noted that for the 6 objects listed, at least one emission line ratio is provided. Additionally, we also classified their excitation class (p) using the method outlined in \cite{1991Ap&SS.181...73G}. This method is based on various emission line ratios, including [O III]/H$_{\beta}$ and [O III]/He II, Table 1 of their work defines three primary excitation classes: low (1-3), medium (4-8), and high (9-12$^{+}$).
 
\begin{figure*}
\begin{center}
\includegraphics[angle=0,scale=0.5]{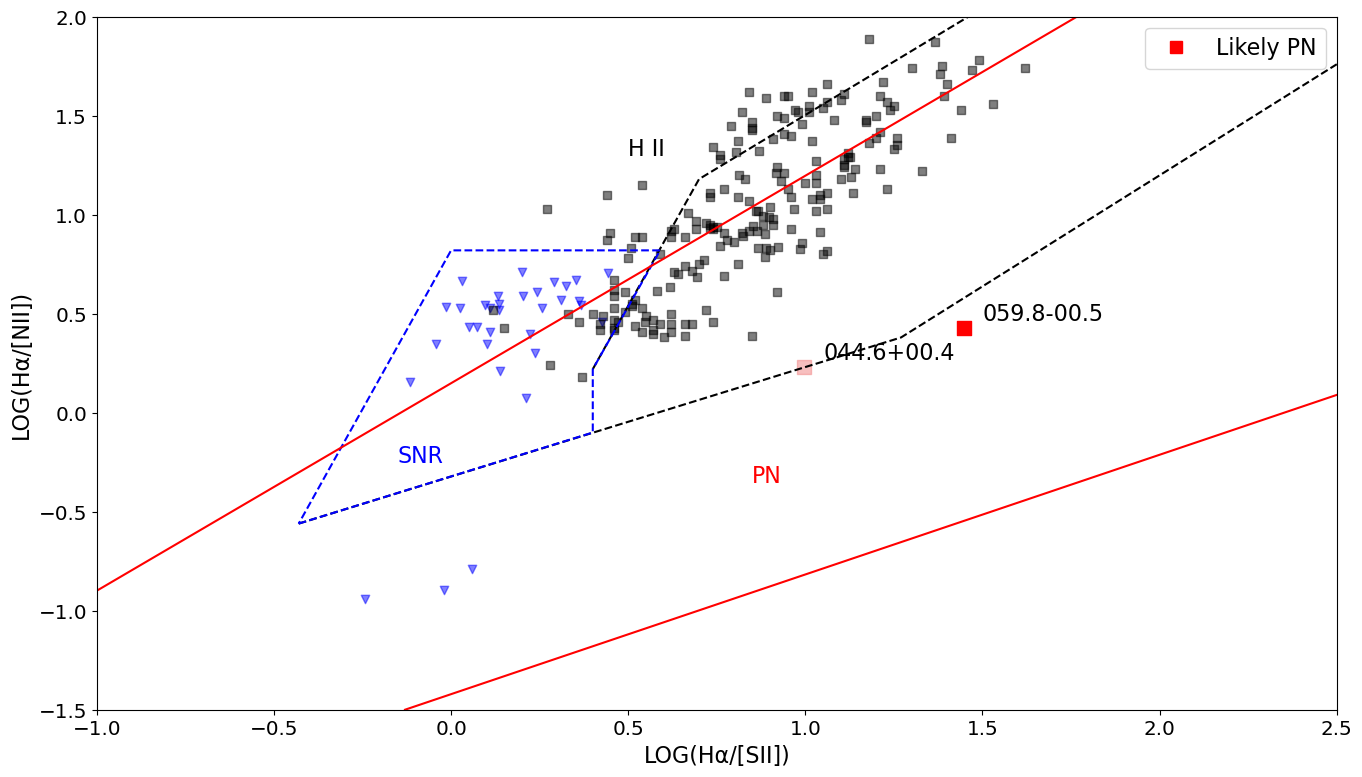}
\caption{SMB diagnostic diagram is constructed using True PN, Likely PN, H II regions, and SNRs. PNe, H II regions (\protect\citealp{2007MNRAS.381.1719V} and references therein) and SNRs (\protect\citealp{2008MNRAS.383.1175P}, and references therein) are represented by a red square, black squares, and blue triangles, respectively. The H II region field is limited by the black dashed lines, while Galactic PNe is located between the red solid lines. SNRs are confined within the blue dashed lines. These restricted regions were determined by \protect\cite{2013MNRAS.431..279S}. The flux ratios of PN G044.6+00.4 (faint red square) were calculated from observed flux without reddening correction.}
\label{F:smb}
\end{center}
\end{figure*}

Emission-line diagnostic diagrams are commonly used in the literature \citep{2010PASA...27..129F, 2013MNRAS.431..279S, 2022MNRAS.512.2202A}. Two primary diagnostic diagrams are the Sabbadin–Minello–Bianchini diagram (SMB) \citep{1977A&A....60..147S} and the Baldwin-Phillips-Terlevich diagram (BPT) \citep{1981PASP...93....5B}.
The SMB diagram uses the log F(H$_{\alpha}$)/F[N II] versus the log F(H$_{\alpha}$)/F[S II] plot. Here F(X) represents the flux of the emission line X, while F[N II] and F[S II] correspond to the total flux of emission lines at 6548+6584 {\AA} and 6716+6731 {\AA} respectively. The BPT diagrams include log (F(6584)/F(H$_{\alpha}$)) versus log (F(5007)/H$_{\beta}$) (BPTa) and log (F([S II]])/F(H$_{\alpha}$)) versus log (F(5007)/H$_{\beta}$) (BPTb). We plotted SMB and BPT diagrams for objects that exhibit these ratios as given in Fig. \ref{F:smb} and Fig. \ref{F:bpt}.

\subsection{Physical Conditions}

Electron temperature ({\it T$_e$}) and electron density ({\it N$_e$}) are used to determine the physical conditions of gaseous nebulae. {\it T$_e$} and {\it N$_e$} of PNe are determined through the analysis of emission line intensities, as certain emission lines are highly sensitive to these physical conditions. Although there are several methods to derive these properties, in this study, we utilized the emission lines of oxygen and nitrogen triplets and sulfur doublet to estimate both {\it T$_e$} and {\it N$_e$}, as they are interdependent parameters. The {\it temden} task in {\scshape IRAF} was used to simultaneously calculate {\it T$_e$} and {\it N$_e$}.

Electron temperature ({\it T$_e$}) is calculated with [O III](${\lambda}$${\lambda}$4959+5007)/${\lambda}$4363 or [N II] ${\lambda}$5755/(${\lambda}$${\lambda}$6548 + 6584) ratios as defined by \cite{2006agna.book.....O}. In the calculation we assumed {\it N$_e$} = 10$^{3}$ cm$^{-3}$. {\it N$_e$} is found using [S II] ${\lambda}$${\lambda}$6717/6731 ratio, assuming {\it T$_e$} = 10$^{4}$ K as given by \cite{2006agna.book.....O}. Here, the lines that have the same excitation energy of [S II] doublets are in use.

\subsection{Ionic Abundances}
 
Studying the abundance of the PNe is crucial for understanding the elemental composition of the interstellar medium (ISM). Therefore, we also determined all the possible ionic abundances for our dataset. We have used PyNeb \citep{2015A&A...573A..42L} for calculating ionic abundances based on 12 + log(n(X$^i$)/n(H$^+$)), where n(X$^i$) is the density of ion that we want to calculate the ionic abundance and n(H$^+$) is the ionic hydrogen density. Temperatures and densities are required for the calculation of abundances in PyNeb. The ionic abundances for objects lacking temperature and density values were calculated assuming {\it N$_e$} = 10$^3$ cm$^{-3}$ and {\it T$_e$} = 10$^4$ K, in accordance with the CASE B conditions outlined by Osterbrock and Ferland (2006)
\cite{2006agna.book.....O}. The ionic abundances of objects are given in Table \ref{T:abd}.

\section{Results and Discussions}
\label{sec:res}
We conducted research on the classification and physical properties of 10 objects by analyzing spectroscopic observations obtained from the TFOSC instrument mounted on the RTT150 telescope. The observations spanned over a period of 4 years, from 2017 to 2020. We have specifically examined the medium-resolution spectra of these 10 objects (given in the Appendix). The measurements of these emission lines, obtained from the spectra, are presented in Table \ref{T:flux}.

4 objects in this study had previously been classified as a Likely PNe (PN G059.8-00.5), including the new candidates (PN G117.2+02.6 and PN G201.5-01.6), and an object of unknown nature (PN G130.4+00.4). All these objects were spectroscopically observed for the first time.

Since one of the most prominent emissions for PNe is the H$_{\alpha}$ line, we have determined the H$_{\alpha}$ fluxes of objects in our list, falling in the range of 6.34x10$^{-9}$ erg cm$^{-2}$ s$^{-1}$ \AA$^{-1}$ (PN G059.8-00.5) to 9.97x10$^{-15}$ erg cm$^{-2}$ s$^{-1}$ \AA$^{-1}$ (PN G117.2+02.6). These H$_{\alpha}$ flux values align with the typical range of H$_{\alpha}$ fluxes observed in Galactic PNe, as reported by \cite{2013MNRAS.431....2F}.

\begin{figure*}
\begin{center}
\includegraphics[angle=0,scale=0.5]{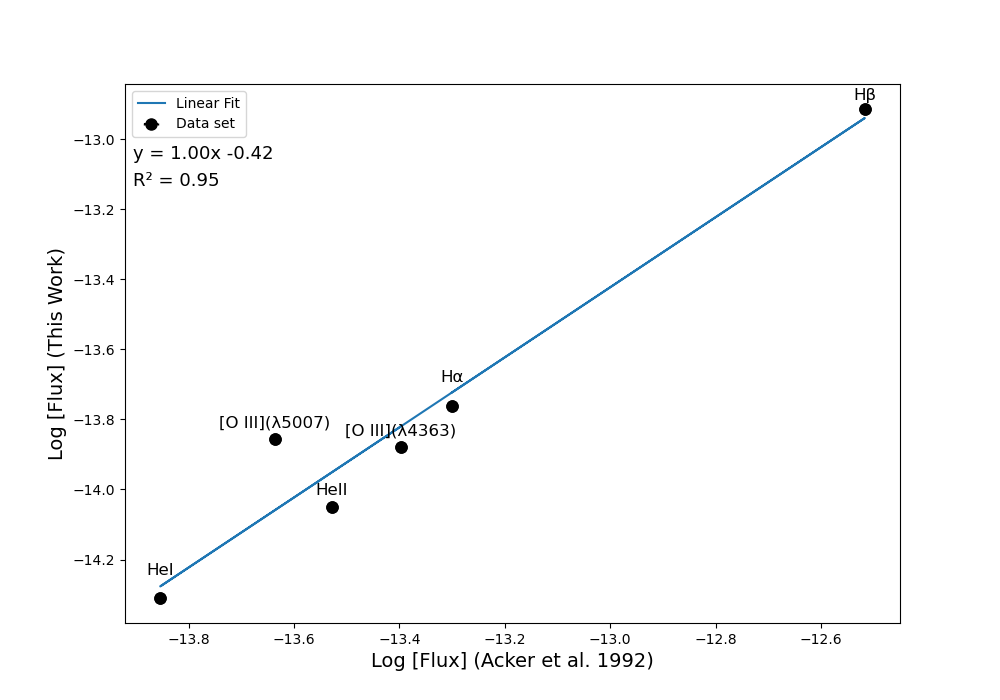}
\caption{The comparison of emission line fluxes between our measurements (vertical axis) and the corresponding fluxes from the \protect\cite{1992secg.book.....A} (horizontal axis). The black dots represent the 6 emission line flux values of PN G076.4+01.8. The size of the error bars is smaller than the symbol size. A blue line represents the best-fit linear regression (R$^{2}$ = 0.95) with an equation of y = 1.00x-0.42.}
\label{F:flux_comp}
\end{center}
\end{figure*}

For comparison purposes, we also found a relevant study by \cite{1992secg.book.....A} that provided emission line fluxes for PN G076.4+01.8. A total of six unreddened flux values for this object are plotted in Fig. \ref{F:flux_comp}. A notable correlation was observed between \cite{1992secg.book.....A} and our measurements, as indicated by a high R$^{2}$ value of 0.95 and a linear equation y = 1.00x-0.42.

\subsection{Classification}
\label{sec:class}
The classification of results using the emission line ratios, and diagnostic diagrams criteria are given in Table \ref{T:plusminus}. The table contains information on the PN status and references for classification from the HASH. The first four rows show that the objects were reclassified from Likely PNe to True PNe. 
The last rows of the table indicate that the previous classifications were retained for six of the objects. 

\begin{figure*}
\begin{center}
\includegraphics[angle=0,scale=0.5]{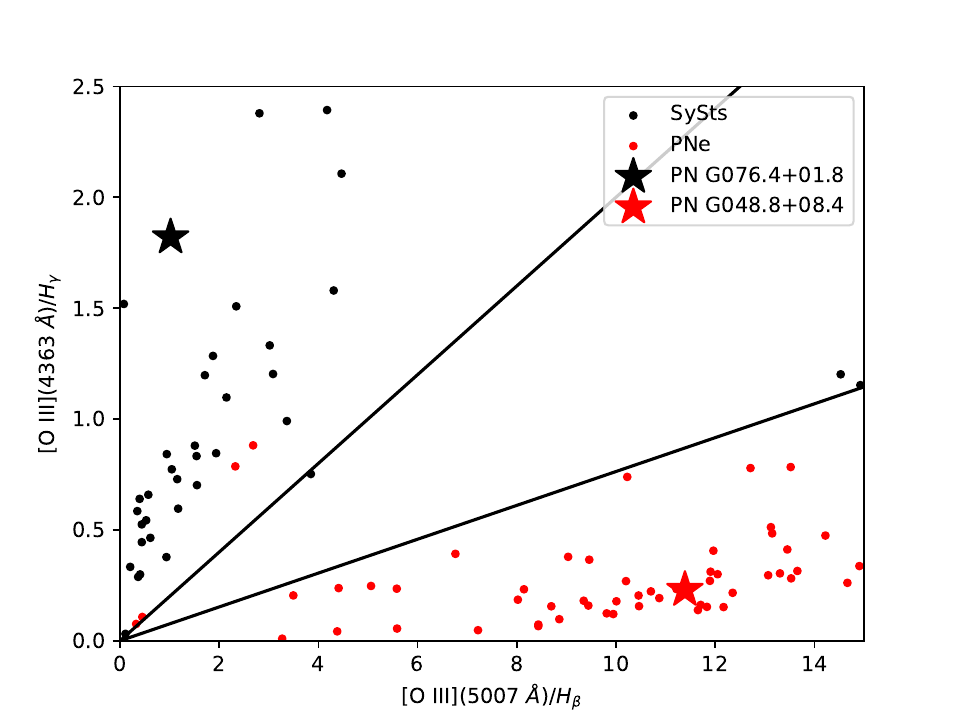}
\caption{[O III]$\lambda$4363/H$_\gamma$ vs. [O III]$\lambda$5007/H$_\beta$ diagnostic diagram of PNe and SySts. The red and black dots correspond to PNe and SySts, respectively. The black and red stars represent the PN G076.4+01.8 and PN G048.8+08.4. The two diagonal lines correspond to an [O III] 5007/4363 line ratio of 13.1 (upper line) and 27.4 (lower line). The data points were taken from the study of \protect\cite{2015A&A...582A..60C}}
\label{F:symbiotic}
\end{center}
\end{figure*}

As seen from Table \ref{T:plusminus}, PN G076.4+01.8 has He II $\lambda$4686 {\AA} line which is a notable diagnostic feature of PNe in their spectra. However, the same He II line is observed in the Symbiotic Stars (SySts) spectra. To discriminate this object we plot [O III] $\lambda$4363/H$_\gamma$ vs. [O III]$\lambda$5007/H$_\beta$ diagnostic diagram, as in \cite{1995PASP..107..462G} and \cite{1989A&A...211L..31S, 2015A&A...582A..60C}, is given in Fig. \ref{F:symbiotic}. As seen in the figure the PN G076.4+01.8 is classified as a possible SySt by this diagnostic diagram. We also plotted the PN G048.8+08.4 in Fig. \ref{F:symbiotic} since it has similar emission lines. However,
the presence of the O VI Raman-scattered emission lines at $\lambda$6825 and $\lambda$7082 {\AA} has also been given as the reliable criterion for identifying symbiotic stars \citep{1989A&A...211L..31S, 2017A&A...606A.110I}. Nevertheless, O VI Raman-scattered lines ($\lambda$6825, $\lambda$7082 {\AA}) could not be detected in the PN G076.4+01.8 spectrum. On the other hand, \cite{2019ApJS..240...21A} reported that approximately \%50 of SySts do not exhibit these specific lines in their spectra. Additionally, \citep{2019MNRAS.483.5077A, 2019MNRAS.488.3238A} provided near-infrared color comparisons for PN G076.4+01.8 with selection criteria for both PNe and SySts. The near-infrared magnitudes for PN G076.4+01.8, obtained from 2MASS \citep{2006AJ....131.1163S}, are as follows: J:15.7, H:15.1, K:14.1, and from WISE \citep{2010AJ....140.1868W}: W1:12.8, W2:11.7, W3:8.3, and W4:6.2. It is important to note that most of these selection criteria do not provide a conclusive determination regarding whether this source should be classified as a PN or a SySts. Consequently, PN G076.4+01.8 remains designated as a Likely PN due to the lack of sufficient evidence for its classification.



\subsubsection{Emission Line Ratios}

Emission line ratios have been used for the determination of emission mechanisms and diagnosis of the objects. Calculated diagnostic line ratios given in Table \ref{T:Test} are [O III]/H$_{\gamma}$, [O III]/H$_{\beta}$, [N II]/H$_{\alpha}$ and [S II]/H$_{\alpha}$. 

In line with the findings of \cite{2010PASA...27..129F}, the [O III]/H$_{\beta}$ ratio is considered more reliable than other emission line ratios. Therefore, based on this ratio, the object PN G021.2+02.9 was classified as a True PN without appearing in diagnostic diagrams due to the absence of required emission line ratios. 3 of 10 objects (PN G021.2+02.9, PN G048.8+08.4, and PN G059.8-00.5) have [O III]/H$_{\beta}$ > 5 emission line ratios as expected from PNe.
Two of them, PN G048.8+08.4 and PN G059.8- 00.5, were classified as likely PNe in the HASH database (see Table \ref{T:plusminus}). For this case, SMB and BPT diagnostic diagrams are needed to confirm whether they are True PNe. We discussed this issue in the next section.

PN G044.6+00.4, PN G048.8+08.4, PN G059.8-00.5, PN G071.4-01.9, and PN G201.5-01.6 can be classified as PNe or H II regions since the ratio of [N II]/H$_{\alpha}$ < 0.6 (see Table \ref{T:plusminus}). On the other hand, the ratio of [S II]/H$_{\alpha}$ < 0.5 indicates that 2 objects (PN G044.6+00.4 and PN G059.8-00.5) could be classified as PNe or H II regions \citep{1985ApJ...292...29F}. However, we could not classify any object according to the emission line ratios of [N II]/H$_{\alpha}$ and [S II]/H$_{\alpha}$. 

\subsubsection{Diagnostic Diagrams}
\label{sec:diag}
SMB and BPT tools play significant roles in diagnosing objects as a PN, an H II region, or an SNR. 2 objects in our list (PN G044.6+00.4 and PN G059.8-00.5) are located within the PN Zone in the SMB diagram. On the other hand, these two objects are also located at the intersection of the H II and PNe zones. However, they are distinct from the area where the concentration of H II regions is prominent, as shown in Fig. \ref{F:smb}. We note that PN G044.6+00.4 is given with observed flux without reddening correction.

\begin{figure*}
\begin{center}
\begin{tabular}{cc}
    \includegraphics[angle=0,scale=0.5]{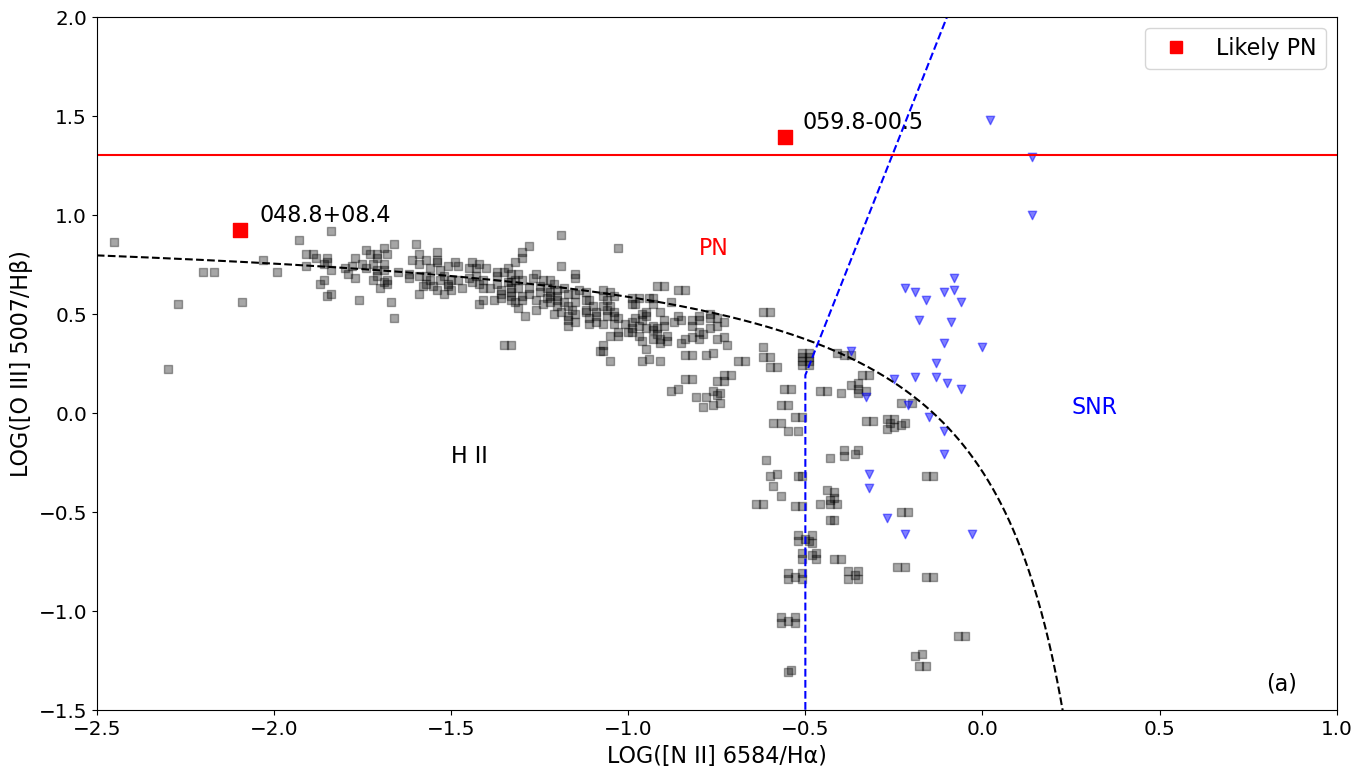}\\
    \includegraphics[angle=0,scale=0.5]{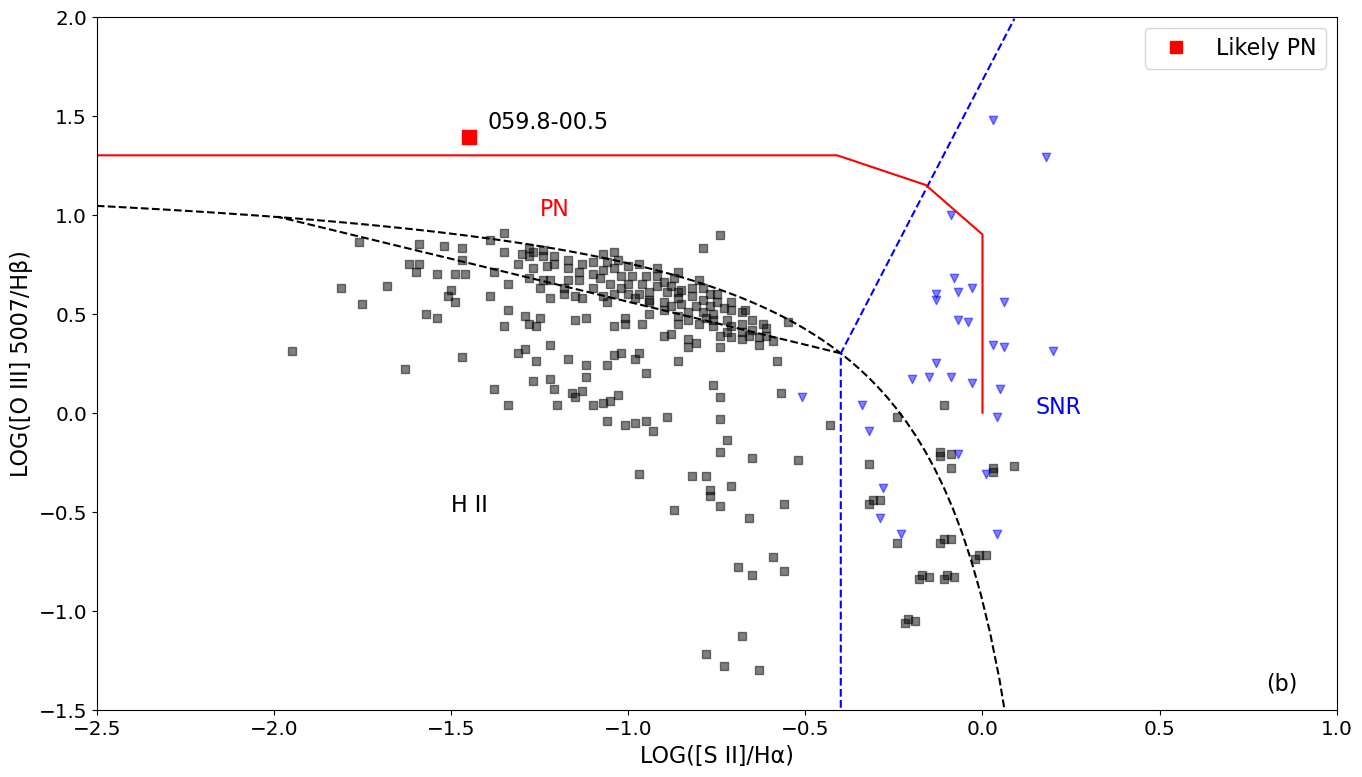}\\
\end{tabular}
\caption{BPTa (a) and BPTb (b) diagram of PNe, H II regions and SNRs. Definitions of the figures are the same as Fig. \ref{F:smb}.}
\label{F:bpt}
\end{center}
\end{figure*}

In addition, 2 of the 10 objects (PN G048.8+08.4 and PN G059.8-00.5) are shown in BPTa PN zone since they have [N II] $\lambda$6584 {\AA} emission line in their spectra (see Fig. \ref{F:bpt}a). 
Furthermore, only one object (PN G059.8-00.5) is also shown in the BPTb diagram (see Fig. \ref{F:bpt}b). The object is located in the PN zone. 

When we examined the BPT diagrams, one object (PN G059.8- 00.5) was found to be located within the PN zone in both diagrams. So, PN G059.8-00.5, which was previously classified as Likely PN, has now been reclassified as True PN.

PN G117.2+02.6, PN G130.4+00.4, PN G201.5-01.6, and PN G212.8-03.6 do not meet any of the aforementioned criteria as they only exhibit the H$_{\alpha}$ emission line in their spectra. Consequently, reclassification of these objects is not possible. Additionally, despite PN G071.4-01.9 having [N II]/H$_{\alpha}$ < 0.6, it is not subject to reclassification based on the above considerations.

Based on our analyses using emission line fluxes, ratios of certain lines, and diagnostic diagrams, we have reclassified four objects that were previously categorized as Likely PNe as True PNe. Additionally, the classification of the six objects remains unchanged.

We also calculated the physical conditions ({\it T$_e$}, {\it N$_e$} and c(H$_{\beta}$) of some PNe and the results are given in Table \ref{T:ext}. In this table, at least one of the physical condition parameters is given for the 6 objects. We also calculate the c(H$_{\beta}$) values from the \cite{1992secg.book.....A} and HASH data. The c(H$_{\beta}$) for PN G048.8+08.4 is not correlated with the HASH spectrum. On the other hand, c(H$_{\beta}$) for PN G048.8+08.4 is compatible with the value found by \cite{1992secg.book.....A}.

The last column of Table \ref{T:Test} shows the excitation class of 4 objects. 3 of them are in the low excitation class and one object is high excitation class. The excitation classes are relevant with ages of the nebulae \citep{1991Ap&SS.181...73G}. We noted that according to the excitation classes of our objects, most of them could be defined are quite young PNe. The excitation classes of PNe give hints about their morphology, the surface temperatures of the central stars, and the brightness of the nebulae. The study of \cite{2009ASPC..404..337K} reported that most of the low excitation nebulae also tend to be Bipolar, Asymmetrical, and Irregular. However, considering the main classes given in Table \ref{T:obslog}, it is difficult to evaluate the morphology of our objects according to their study.

We intended to derive elemental abundances from the determined ionic abundances for some PNe because elemental abundances provide essential information about the composition and chemical evolution of these objects. Furthermore, analyzing the abundances allows for the classification of PNe into different types based on their chemical compositions \citep{1978IAUS...76..215P, 2004MNRAS.349..793P,2010ApJ...714.1096S,2018ApJ...862...45S}. Elemental abundances are calculated by multiplying the ionic abundance by the Ionization Correction Factor (ICF). However, we encountered difficulties in computing the ICF due to the absence of required emission lines such as [O II] and He I. Therefore, we are unable to access the mentioned information regarding the abundance based on our findings. In addition, our objects in Table \ref{T:abd} are classified as low ionization PNe. As indicated by \cite{2013A&A...558A.122G}, the absence of a He II emission line within the spectrum indicates a state of low ionization. When we analyze our spectra, it is clear that the majority of objects do not exhibit He II emission. After a comparison between our computed ionic abundances and the results presented in \cite{2013A&A...558A.122G}, we noticed that the ionic abundances for Cn 1-5, He 2-86, M 1-30, M 1-61, and M 3-15 are similar.

\section{Conclusions}
\label{sec:con}

We investigated spectroscopically 6 Likely PNe, 2 new candidates, one emission line star, and one object of unknown nature. These are selected from HASH and are characterized by a small angular size (< 8\arcsec) and a location in the northern hemisphere. We have classified the objects through the analysis of their emission line ratios, and using diagnostic diagrams such as SMB and BPT. In addition to the classification of these objects, we also determined their physical parameters and ionic abundances. Our main findings can be summarized as follows:

\begin{enumerate}
\item Four objects in our sample were observed for the first time with a medium-resolution spectrograph (TFOSC). These 4 objects are denoted in HASH as Likely PN (PN G059.8-00.5), new candidates (PN G117.2+02.6 and PN G201.5-01.6), and unknown nature (PN G130.4+00.4). In this study, PN G059.8-00.5 is reclassified as True PN. The status of the remaining three objects did not change due to the absence of required emission lines. 

\item  
4 Likely PNe of 10 (PN G021.2+02.9, PN G044.6+00.4, PN G048.8+08.4, and PN G059.8-00.5) are reclassified as True PNe. 

\item
PN G076.4+01.8 retains its Likely PN classification due to the absence of O VI Raman-scattered emission lines at $\lambda$6825 and $\lambda$7082 {\AA}, which are typical features of almost half of SySts. Additionally, we do not have a strong level of confidence in the provided near-infrared selection criteria for SySts.

\item Furthermore, the ionic abundances that we have found, are compatible with the abundances expected from low ionization PNe \citep{2013A&A...558A.122G}.

\item Three objects (PN G021.2+02.9, PN G048.8+08.4, and PN G059.8-00.5) in our list belong to the low excitation class and exhibit elliptical or spherical morphology. We noted that these findings are inconsistent with \cite{2009ASPC..404..337K} interpretation of object morphology.

The spectra of the 10 objects will contribute to the update of the HASH database, and the ongoing project focuses on the confirmation of the objects in HASH as True PNe.

\end{enumerate}

\section*{Acknowledgements}
This research was supported by the Scientific and Technological Research Council of Turkey (TÜBİTAK) through project number 122F122. The authors thank the TÜBİTAK for partial support in using RTT150 (Russian–Turkish 1.5-m telescope) in Antalya through project number 16BRTT150-1064 and 22ARTT150-1925. We also thank TÜBİTAK National Observatory (TUG) staff. We special thanks to N. Erzincan for the help with the visualization of graphics. This work has made use of data from the European Space Agency (ESA) mission Gaia (https://www.cosmos.esa.int/gaia), processed by the Gaia Data Processing and Analysis Consortium (DPAC, https://www.cosmos.esa.int/web/gaia/dpac/consortium). Funding for the DPAC has been provided by national institutions, in particular, the institutions participating in the Gaia Multilateral Agreement

\section*{Data Availability}

The data may be available from the corresponding author on request.


\bibliographystyle{mnras}
\bibliography{pne} 

\section{APPENDIX A: PNe SPECTRA}
In this appendix, we provide figures of spectra for all PNe observations divided into 2 subsections. Fig. \ref{F:A1} presents spectra for the 3 newly classified True PNe and one Likely PN; Fig. \ref{F:A2} shows new spectra of previously observed objects. The figures provide the IAU PN G designation.

\begin{table*}
\centering
\caption{The list of selected objects from the HASH is shown. They are ordered by the PN G number. The columns are represented the catalog number, name of the object, right ascension, declination, the major diameter of objects, main class and PN status from \href{http://202.189.117.101:8999/gpne/index.php}{HASH}, R$_{mag}$ magnitudes from {\it USNO B1.0} catalog \protect\citep{2003AJ....125..984M}, distances from {\it GAIA} DR3 \protect\citep{2016A&A...595A...1G, 2022arXiv220800211G}, Observation Dates, Grisms and Seeing. Seeing is calculated from the mean point spread function of field stars in the images taken before the object spectrum. Bold faces indicate objects observed spectroscopically for the first time.}
    \begin{tabular}{@{}l@{~~}c@{~~}c@{~~}c@{~~}c@{~~}c@{~~}c@{~~}c@{~~}c@{~~}c@{~~}c@{~~}r@{}}
    \hline\hline
    PN G & Name & R.A (J2000) & Dec. (J2000) & Major Diam. & Main Class$^{\dagger}$ & PN Status$^{\dagger\dagger}$ & R$_{mag}$ & Distance & Obs Date & Grisms  & Seeing \\
    &&(hh:mm:ss.s)& (dd:mm:ss.s) & (\arcsec) & & &  & (kpc) & (yyyymmdd) &  & (\arcsec) \\
\hline
 021.2+02.9 	&	\mbox{J1819-0901}$^{a}$	&	18:19:06.7	&	-09:01:50.3	&	7	&	 E 	&	 L 	&	16.8	&	1.3	&	20180713	&	 8 - 14 	&	 1.8 \\
 044.6+00.4 	&	\mbox{19089+1032}$^{b}$	&	19:11:17.2	&	10:37:34.2	&	4.2	&	 S 	&	 L 	&	16.8	&	-	&	20180714	&	 8 - 14 	&	 1.8 \\
 048.8+08.4 	&	\mbox{Kn 66}	&	18:49:54.9	&	17:57:15.0	&	2	&	 S 	&	 L 	&	15.9	&	1.2	&	20190728	&	 8 - 14 	&	 1.8 \\
 \textbf{059.8-00.5} 	&	\mbox{J19453289+2328105}$^{c}$	&	19:45:32.9	&	23:28:10.5	&	2.5	&	 E 	&	 L 	&	15.5	&	1.3	&	20200516	&	 8 - 14 	&	 1.5 \\
 071.4-01.9 	&	\mbox{PM 1-323}	&	20:18:50.0	&	32:36:23.9	&	7	&	 B 	&	 L 	&	18.7	&	-	&	20180713	&	 8 - 14 	&	 1.8 \\
 076.4+01.8 	&	\mbox{KjPn 3}	&	20:17:15.5	&	38:50:23.9	&	3	&	 S 	&	 L 	&	15.9	&	2.6	&	20200517	&	 8 - 14 	&	  - \\
 \textbf{117.2+02.6} 	&	\mbox{23545+6441}$^{b}$	&	23:57:00.2	&	64:57:25.1	&	 - 	&	 - 	&	 c 	&	19.0	&	1.7	&	20170131	&	 8 - 14 	&	 3.6 \\
 \textbf{130.4+00.4} 	&	\mbox{J015658.96+622406.8}$^{d}$	&	01:56:59.0	&	62:24:06.8	&	 - 	&	 - 	&	 ooun 	&	 - 	&	-	&	20170131	&	 8 - 14 	&	 1.8 \\
 \textbf{201.5-01.6} 	&	\mbox{J062438.77+091825.9}$^{d}$	&	06:24:38.8	&	09:18:25.9	&	 - 	&	 - 	&	 c 	&	17.5	&	0.8	&	20170129	&	 8 - 14 	&	 1.8 \\
 212.8-03.6 	&	\mbox{06357-0130}$^{b}$	&	06:38:16.1	&	-01:33:08.8	&	 - 	&	 S 	&	 Em* 	&	14.3	&	-	&	20170129	&	 8 - 14 	&	 1.7 \\
    \hline
    \end{tabular}
    \\
        $^{a}$:MPA, $^{b}$:IRAS,$^{c}$:2MASS, $^{d}$:IPHAS, $^{e}$:IPHASX \\
       $^{\dagger}$ E: Elliptical, S: Spherical, B: Bipolar according to the ERBIAS scheme in \cite{2022FrASS...9.5287P}\\
       $^{\dagger\dagger}$ T: True PN, L: Likely PN, P: Possible PN, c: New Candidate, ooun: Object of Unknown Nature, Em*: Emission Line Star.
\label{T:obslog}
\end{table*}

\begin{table*}
  \caption{The reddening corrected emission line fluxes of the objects included in this study are listed. All line fluxes are scaled to their H$_{\beta}$ flux (= 100). The {\it splot} task calculates error values of fluxes. Some lines denoted as "-" are not detected in their spectra. The measured flux values for objects marked with an asterisk (*) have been scaled with respect to H$_{\alpha}$ and not reddening corrected due to the lack of H$_{\beta}$ emission line.}
    \begin{turn}{90}
    \begin{tabular}{c@{~}c@{~}c@{~}c@{~}c@{~}c@{~}c@{~}c@{~}c@{~}c@{~}c@{~}c@{~}c@{~}c@{~}}
    \hline\hline
    PN G & [O III] & He II & [O III] & [O III] & [O I] & [N II] & H$_{\alpha}$& [N II] & [S II] & [S II] & He I & [Ar III] \\
       &	$\lambda$4363&$\lambda$4686&$\lambda$4959&$\lambda$5007&$\lambda$6300&$\lambda$6548&$\lambda$6563&$\lambda$6584&$\lambda$6716&$\lambda$6731&$\lambda$7065&$\lambda$7136 \\
     \hline
021.2+02.9	&		-		&		-		&	216.46	$\pm$	14.91	&	625.20	$\pm$	14.23	&		-		&		-		&	330.22	$\pm$	1.08	&		-		&		-		&		-		&	6.23	$\pm$	1.49	&	11.63	$\pm$	1.36	\\
044.6+00.4$^{*}$	&		-		&		-		&	12.86	$\pm$	2.09	&	28.10	$\pm$	2.00	&		-		&	16.16	$\pm$	0.76	&	100.00	$\pm$	0.63	&	42.95	$\pm$	0.61	&	3.58	$\pm$	0.54	&	6.50	$\pm$	0.74	&	6.74	$\pm$	0.59	&	16.50	$\pm$	0.67	\\
048.8+08.4	&	17.99	$\pm$	0.75	&		-		&	297.33	$\pm$	0.81	&	842.04	$\pm$	0.81	&		-		&		-		&	343.84	$\pm$	0.30	&	2.76	$\pm$	0.38	&		-		&		-		&	4.50	$\pm$	0.23	&	9.38	$\pm$	0.29	\\
059.8-00.5	&		-		&		-		&	900.66	$\pm$	0.99	&	2483.44	$\pm$	1.11	&	10.61	$\pm$	0.94	&	36.71	$\pm$	0.82	&	371.02	$\pm$	0.82	&	102.60	$\pm$	0.88	&	4.29	$\pm$	1.05	&	8.92	$\pm$	0.94	&	8.86	$\pm$	0.94	&	20.52	$\pm$	0.76	\\
071.4-01.9$^{*}$ &		-		&		-		&		-		&	12.57	$\pm$	0.95	&		-		&		-		&	100.00	$\pm$	0.92	&	6.34	$\pm$	0.78	&		-		&		-		&		-		&		-		\\
076.4+01.8	&	112.51	$\pm$	1.17	&	61.68	$\pm$	0.99	&	27.39	$\pm$	1.11	&	74.86	$\pm$	0.92	&	5.12	$\pm$	0.33	&		-		&	349.29	$\pm$	0.36	&		-		&		-		&		-		&	13.58	$\pm$	0.29	&	4.01	$\pm$	0.33	\\
117.2+02.6	&		-		&		-		&		-		&		-		&	47.67	$\pm$	2.67	&		-		&	268.90	$\pm$	3.51	&		-		&		-		&		-		&		-		&		-		\\
130.4+00.4$^{*}$ &		-		&		-		&		-		&		-		&		-		&		-		&	100.00	$\pm$	0.48	&		-		&		-		&		-		&				&		-		\\
201.5-01.6	&		-		&		-		&		-		&		-		&	9.39	$\pm$	1.80	&	44.72	$\pm$	2.22	&	509.61	$\pm$	2.10	&		-		&		-		&		-		&		-		&		-		\\
212.8-03.6	&		-		&		-		&		-		&		-		&		-		&		-		&	464.56	$\pm$	1.39	&		-		&		-		&		-		&		-		&		-		\\
   \hline
    \end{tabular}
 \label{T:flux}
\end{turn}
\end{table*}

\begin{table*}
\caption{Emission line ratios and excitation classes for the majority of the objects. The excitation class has been shown by p \protect\citep{1991Ap&SS.181...73G}. The classes from p are low (1-3), medium (4-8), and high (9-12+). Line ratios are ordered with the wavelength. The errors are propagated from flux values given in Table \ref{T:flux}. In the calculated ratios of objects marked with asterisks "$^{*}$", uncorrected fluxes were utilized as these objects did not exhibit H$_{\beta}$ flux.
}
\begin{tabular}{lcc@{~}ccc}
\hline\hline
PN G	& [O III]/H$_{\gamma}$ & [O III]/H$_{\beta}$	& [N II]/H$_{\alpha}$	& [S II]/H$_{\alpha}$ & p \\
\hline
021.2+02.9	&		-		&	8.417	$\pm$	1.282	&		-		&		-		&	2	\\
044.6+00.4$^{*}$	&		-		&		-		&	0.591	$\pm$	0.013	&	0.101	$\pm$	0.012	&	-	\\
048.8+08.4	&	0.234	$\pm$	0.010	&	11.394	$\pm$	0.086	&	0.008	$\pm$	0.001	&		-		&	3	\\
059.8-00.5	&		-		&	33.841	$\pm$	0.298	&	0.375	$\pm$	0.004	&	0.036	$\pm$	0.005	&	3	\\
071.4-01.9$^{*}$ 	&		-		&		-		&	0.063	$\pm$	0.008	&		-		&	-	\\
076.4+01.8	&	1.824	$\pm$	0.039	&	1.023	$\pm$	0.021	&		-		&		-		&	12+	\\
201.5-01.6	&		-		&		-		&	0.088	$\pm$	0.004	&		-		&	-	\\
\hline
\end{tabular}
\label{T:Test}
\end{table*}

\begin{table*}
\centering
\caption{The available values for electron temperature ({\it T$_e$}), electron density ({\it N$_e$}), and logarithmic extinction constant c(H$_{\beta}$) are shown.
}
\begin{tabular}{lccc}
\hline\hline
PN G & {\it T$_e$} & {\it N$_e$} & c(H$_{\beta}$) \\
&(K)& (cm$^{-3}$)& \\
\hline
021.2+02.9	&	-	&	-	&	1.25	\\
048.8+08.4	&	15759	&	-	&	0.96 / 0.25$^{\dagger}$	\\
059.8-00.5	&	-	&	20736	&	5.80	\\
076.4+01.8	&	-	&	-	&	1.03 / 0.93$^{a}$	\\
201.5-01.6	&	-	&	-	&	3.15	\\
212.8-03.6	&	-	&	-	&	2.38	\\
\hline
\end{tabular}
\\
    $^{a}$ : Values of c(H$_{\beta}$) inferred from H$_{\alpha}$ and H$_{\beta}$ supplied to \cite{1992secg.book.....A} \\
    $^{\dagger}$ : Values of c(H$_{\beta}$) inferred from H$_{\alpha}$ and H$_{\beta}$ supplied to HASH spectrum.
\label{T:ext}
\end{table*}

\begin{table*}
    \caption{Ionic abundances of some objects are calculated by PyNeb. Only objects with H$_{\beta}$ emission line have been tabulated. [O III], [N II] and [S II] values are the mean abundances for [O III] $\lambda$$\lambda$4959, 5007 {\AA}, [N II] $\lambda$$\lambda$6548, 6584 {\AA} and [S II] $\lambda$$\lambda$6716, 6731 {\AA} respectively. Lines are ordered with the wavelength and the table format is 12 + log(n(X$^{i}$)/n(H$^+$)).}
    \begin{tabular}{ccccccc}
    \hline\hline
    PN G	&	O$^{++}$ / H$^{+}$	&	N$^{+}$ / H$^{+}$  	&	S$^{+}$ / H$^{+}$  & Ar$^{++}$ / H$^{+}$ 	\\
    \hline																					
    021.2+02.9	&	8.44	&	-	&	-	&	6.45	\\
    048.8+08.4	&	8.51	&	5.41	&	-	&	5.69	\\
    059.8-00.5	&	8.63	&	7.24	&	5.91	&	8.65	\\
    076.4+01.8	&	7.43	&	-	&	-	&	5.83	\\
    201.5-01.6	&	-	&	8.31	&	-	&	-	\\
    \hline
    \end{tabular}
\label{T:abd}
\end{table*}

\begin{table*}
\centering
\caption{The results of diagnostic criteria and diagrams for objects. The reference to the early classification of objects is taken from HASH. He II criteria were added as the middle column. Emission line ratio criteria were provided. SMB, BPTa, BPTb, and \protect\cite{2015A&A...582A..60C} columns are also provided as the results of diagnostic diagrams. The last column represents the results of the diagnosis. According to our results, the status changes are indicated in bold in the last column.}
    \begin{tabular}{ccccccccccccc}
    \hline\hline
    PN G & PN Satus & Ref. &He II & [O III]/H$_{\beta}$ > 5 & [N II]/H$_{\alpha}$ < 0.6 & [S II]/H$_{\alpha}$ < 0.5 & SMB & BPTa & BPTb &  Clyne et al. (2015) & Class \\
    \hline
021.2+02.9	&	L	&	(1)	&	-	&	PN	&	-	&	-	&	-	&	-	&	-	& -	&	\textbf{T}	\\
044.6+00.4	&	L	&	(2)	&	-	&	-	&	PN/H II	&	PN/H II	&	PN	&	-	&	-	& -	&	\textbf{T}	\\
048.8+08.4	&	L	&	(3)	&	-	&	PN	&	PN/H II	&	-	&	-	&	PN	&	-	& PN &	\textbf{T}	\\
059.8-00.5	&	L	&	(2)	&	-	&	PN	&	PN/H II	&	PN/H II	&	PN	&	PN	&	PN	& -	&	\textbf{T}	\\
    \hline
071.4-01.9	&	L	&	(4) &	-	&	-	&	PN/H II	&	-	&	-	&	-	&	-	&	-	& L	\\
076.4+01.8	&	L	&	(5)	&	PN	&	SNR/H II	&	-	&	-	&	-	&	-	&	-	& SySts	&	L	\\
117.2+02.6	&	c	&	(2)	&	-	&	-	&	-	&	-	&	-	&	-	&	-	& -	&	c	\\
130.4+00.4	&	ooun	&	(2)	&	-	&	-	&	-	&	-	&	-	&	-	&	-	& -	&	ooun	\\
201.5-01.6	&	c	&	(2)	&	-	&	-	&	PN/H II	&	-	&	-	&	-	&	-	& -	&	c	\\
212.8-03.6	&	Em*	&	(2)	&	-	&	-	&	-	&	-	&	-	&	-	&	-	& -	&	Em*	\\
    \hline
    \end{tabular}
    \\
(1):\cite{2008MNRAS.384..525M},  (2):\cite{2009A&A...504..291V}, (3):\cite{2014apn6.confE..48K}, (4):\cite{1988A&AS...76..317P},(5):\cite{1992secg.book.....A}
    \label{T:plusminus}
\end{table*}

\appendix
\setcounter{figure}{0} \renewcommand{\thefigure}{A.\arabic{figure}}

\begin{figure*}
\caption{The flux-calibrated spectra of four reclassified PNe. The first panel shows the spectrum of PN G021.2+02.9, classified as a True PN based on the criterion of [O III]/H$_{\beta}$ > 5. The second panel shows the spectrum of PN G044.6+00.4, which suffers from extinction in the blue part of the spectrum, making the detection of [O III] and H$_{\beta}$ problematic. However, it has separated [N II] emission lines and barely detected [S II] emission lines. This PN is reclassified as a True PN using the diagnostic diagram of the SMB. The third panel shows the spectrum of PN G048.8+08.4, which has a strong [O III] emission line as expected from a PN, as well as an H$_{\alpha}$ line in the red part of the spectrum. The fourth panel shows the spectrum of PN G076.4+01.8, which is a Likely PN (see section \ref{sec:class}). Definitions of the figures are the same as Fig. \ref{F:spec}.}
\centering
\begin{tabular}{@{}c@{}}
\includegraphics[angle=0,width=0.85\textwidth]{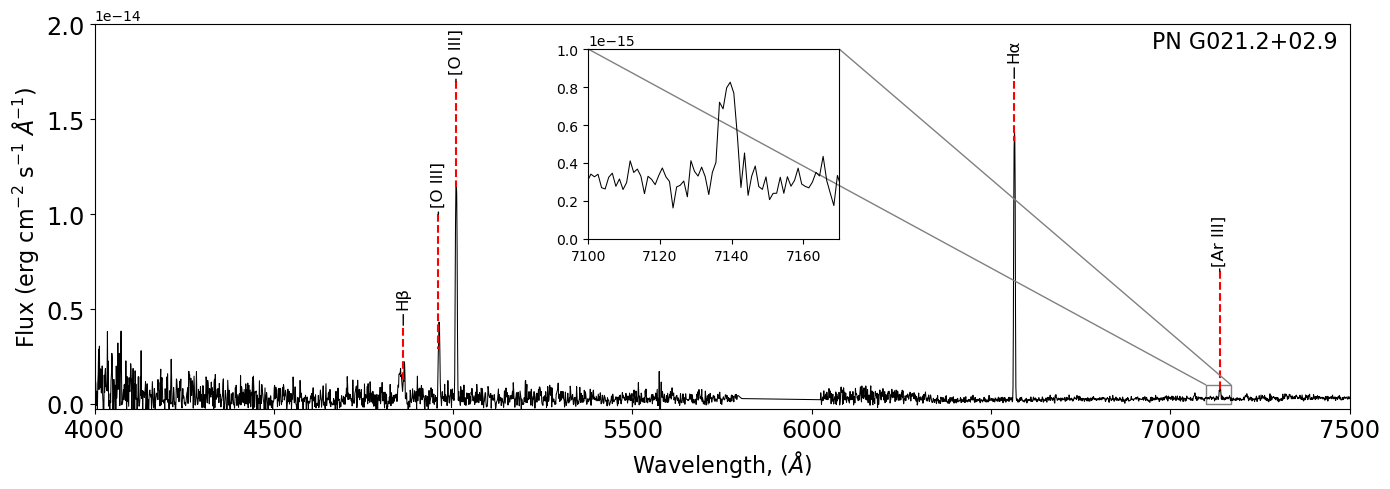}\\
\includegraphics[angle=0,width=0.85\textwidth]{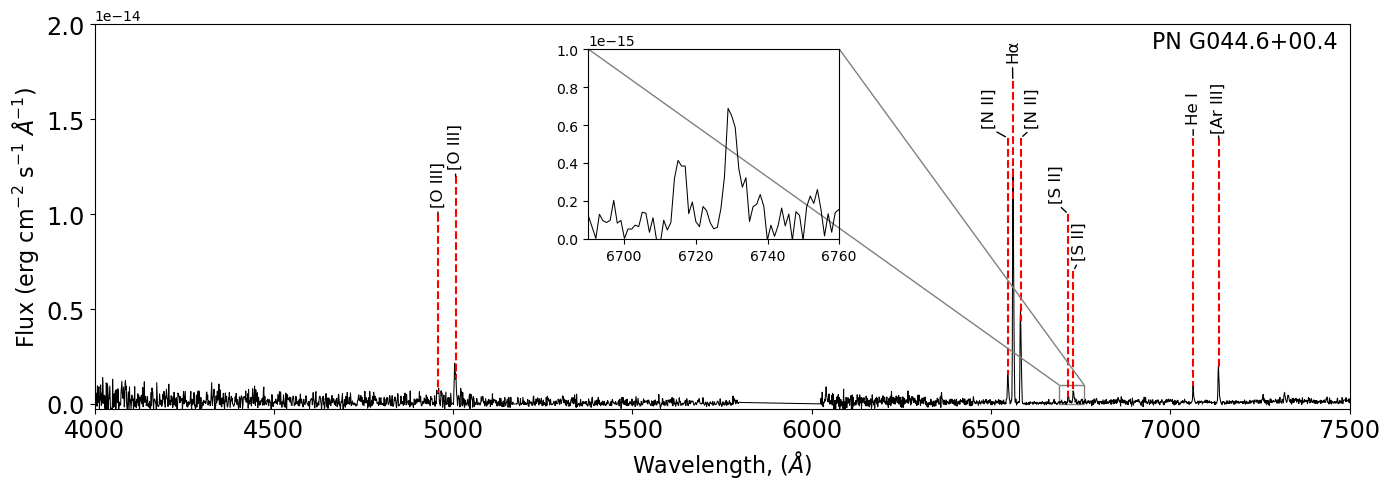}\\
\includegraphics[angle=0,width=0.85\textwidth]{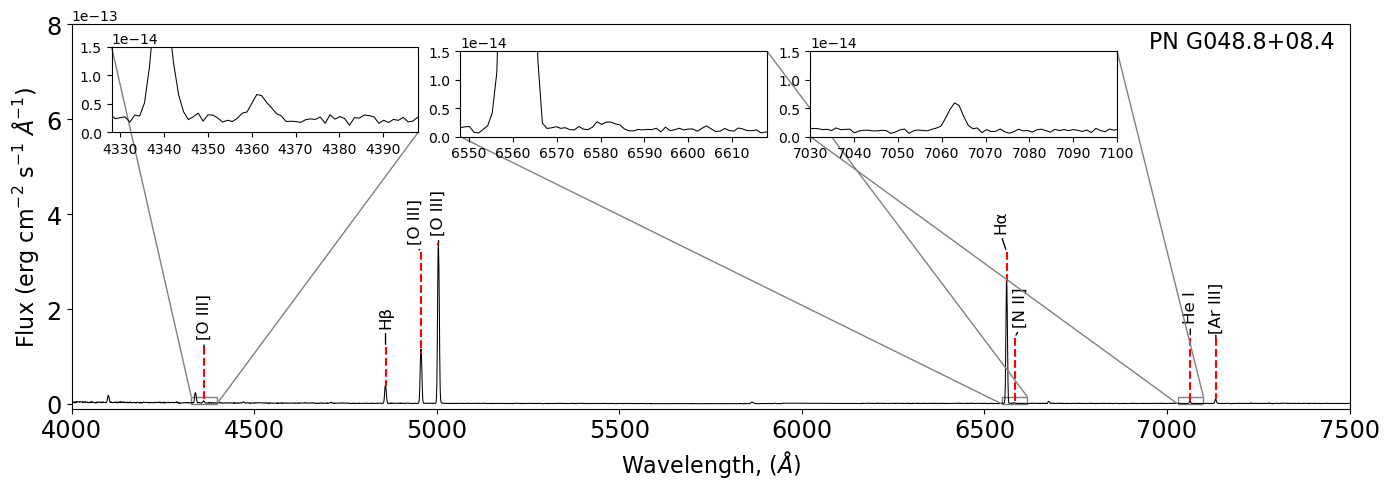}\\
\includegraphics[angle=0,width=0.85\textwidth]{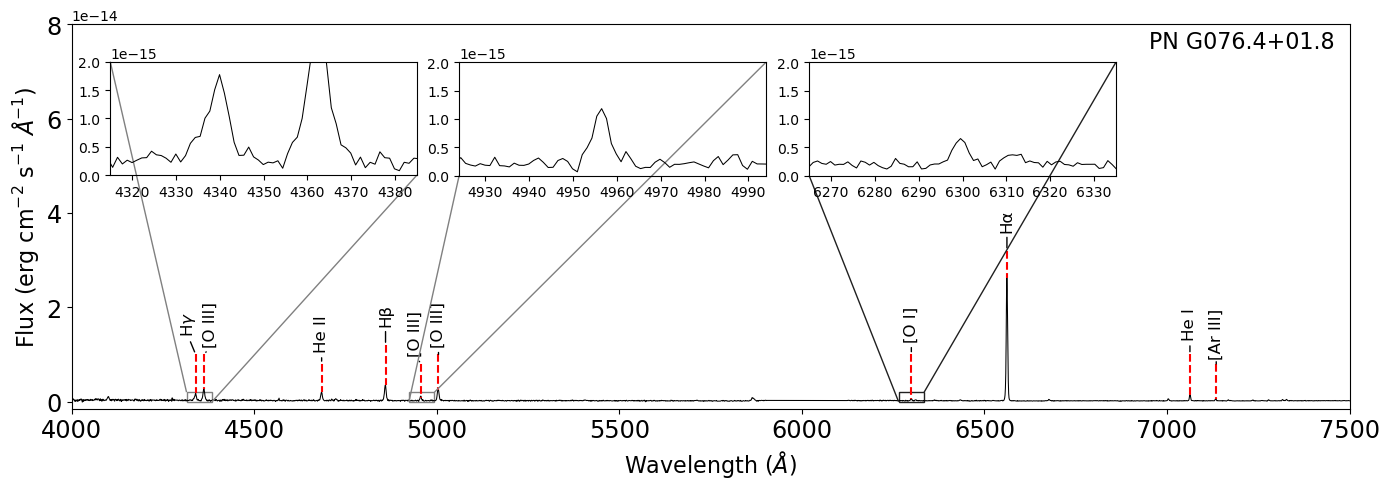}\\
\end{tabular}
\label{F:A1}
\end{figure*}

\begin{figure*}
\caption{The spectra were obtained in this study. Definitions of the figures are the same as Fig. \ref{F:spec}}
\label{F:A2}
\centering
\begin{tabular}{@{}c@{}c@{}}
\includegraphics[angle=0,width=0.49\textwidth]{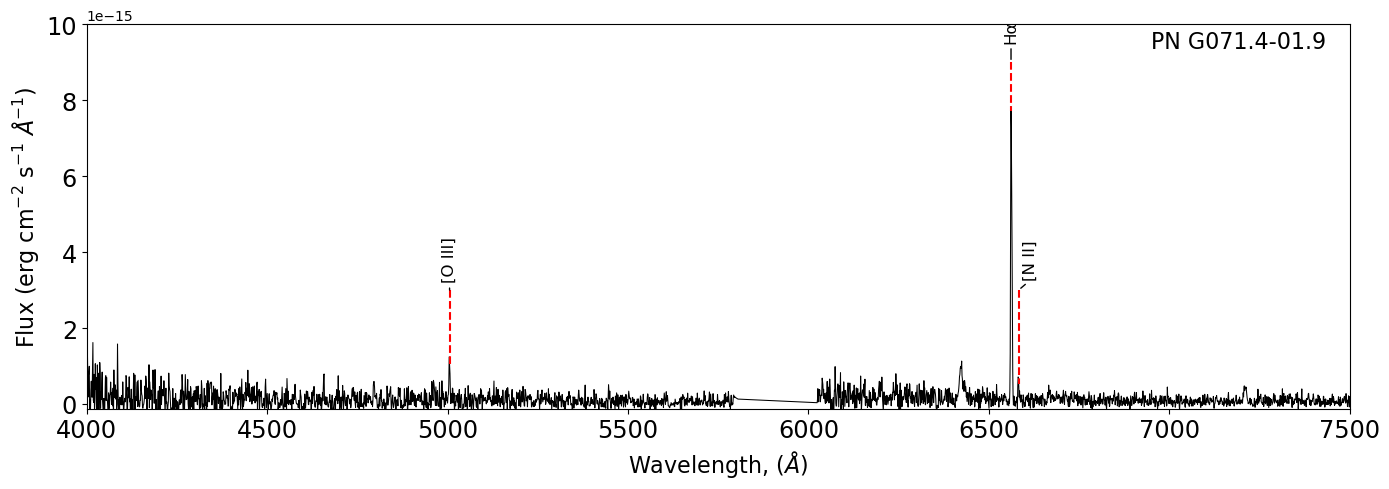}&
\includegraphics[angle=0,width=0.49\textwidth]{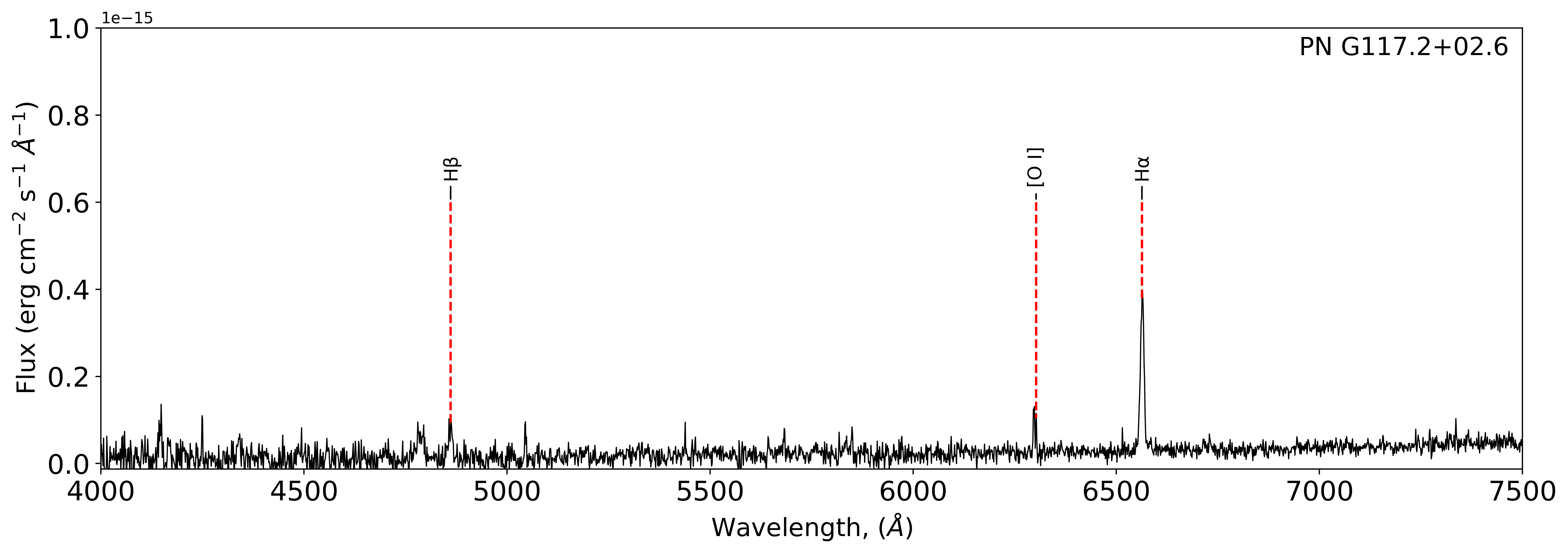}\\
\includegraphics[angle=0,width=0.49\textwidth]{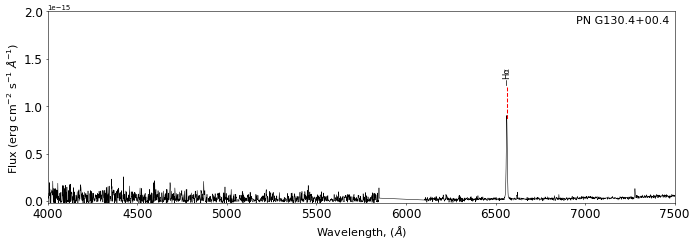}&
\includegraphics[angle=0,width=0.49\textwidth]{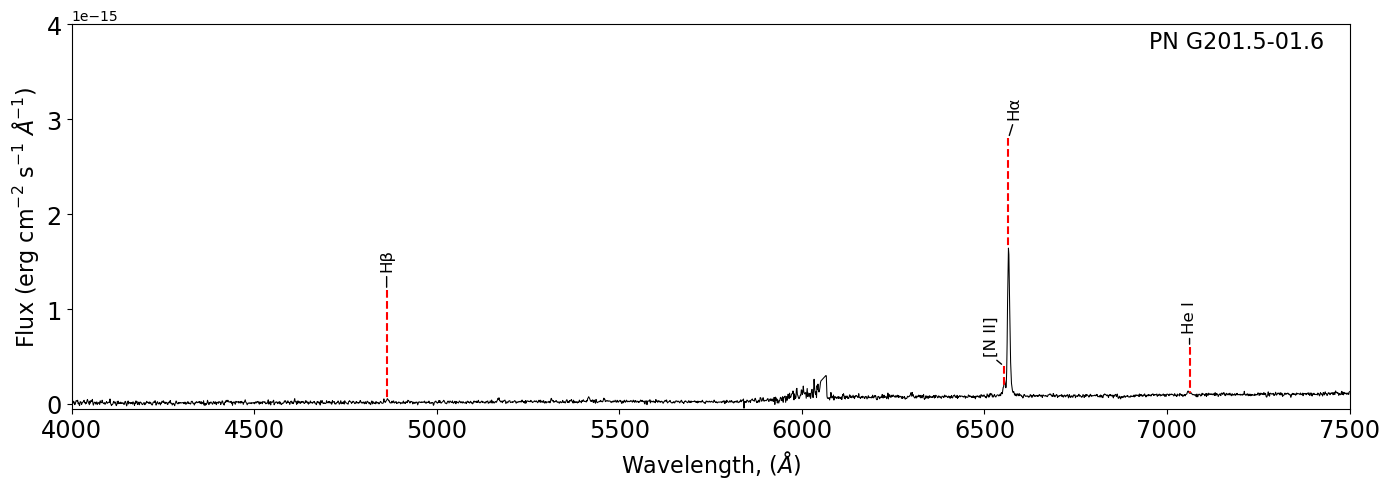}\\
\includegraphics[angle=0,width=0.49\textwidth]{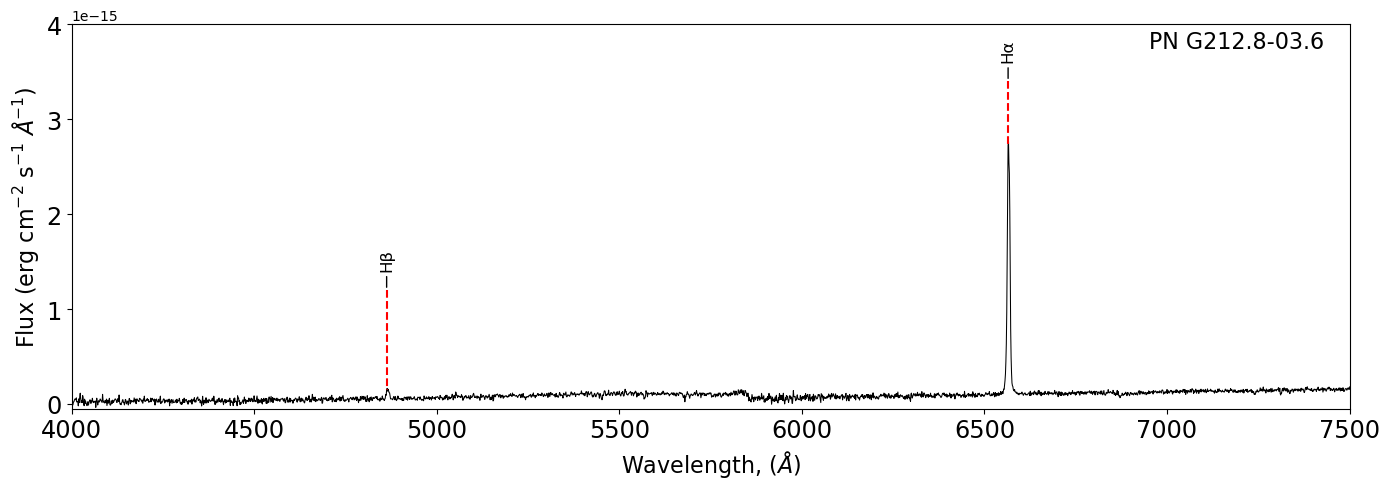}\\
\end{tabular}
\end{figure*}

\bsp	
\label{lastpage}
\end{document}